\documentclass[a4paper,11pt]{article}
\usepackage{pos}

\title{Measuring system-size and event-topology dependence of (multi-)strangeness production}

\author*[a]{Lucia Anna Tarasovi\v{c}ov\'{a}}
\onbehalf{for the ALICE Collaboration}

\affiliation[a]{Pavol Jozef Šafárik University,\\
Šrobárova 2, 041 54 Košice, Slovakia}

\emailAdd{lucia.anna.husova@cern.ch}

\abstract{Measurements of light-flavour particle production in small collision systems at the~LHC energies have shown the~onset of features that resemble what is typically observed in nucleus-nucleus collisions. New results on the~(multi-)strange hadron production in Pb--Pb collisions at $\sqrt{s_{\mathrm{NN}}}$=5.02 and 5.36~TeV will be presented. These results are discussed in the~context of recent measurements of light-flavour hadron production in pp collisions at $\sqrt{s}$~=~0.9 and 13.6~TeV collected by the~ALICE experiment in Run 3 of the LHC. 
In order to understand the~strangeness production mechanism, angular correlation between multi-strange and associated identified hadrons are measured and compared with predictions from the~string-breaking model PYTHIA8, the~cluster hadronisation model HERWIG7, and the~core-corona model EPOS-LHC.
In addition, the connection of strange hadron production to hard scattering processes and to the~underlying event is studied, using di-hadron correlations triggered with the highest-$p_{\mathrm{T}}$ hadron in the event.}

\FullConference{42nd International Conference on High Energy Physics (ICHEP2024)\\
18-24 July 2024\\
Prague, Czech Republic\\}

\begin{document}
\maketitle

\section{Introduction}

Under the~extreme conditions (high temperature and pressure) occurring during heavy-ion collisions, a deconfined state of matter, known as quark--gluon plasma (QGP), can be created. 
One of the~proposed consequences of the presence of the QGP during the collision evolution in the final state is the observation of an enhanced production of strange hadrons over pions. 
However, such a~trend as a function of multiplicity is not only observed in the collisions of heavy ions~\cite{ALICE:2013xmt,STAR:2007cqw}, but also in smaller systems like p--Pb or pp collisions~\cite{ALICE:2016fzo}. 
Moreover, the increase is steeper for hadrons with higher strangeness content.
This is similar in all systems as it would be expected from a thermal production of strangeness.
The origin of this trend in small collision systems is still not clearly explained as pp collisions are generally understood to be too small for any kind of medium, if created, to be thermalised.

One of the possible sources of the enhanced strangeness production could be jet fragmentation or a bias between baryon and meson production. 
In order to review this possibilities, the knowledge of the hadronisation mechanism is crucial. 
Nevertheless, hadron production is implemented in different ways in the models which are used to describe the data. 
The PYTHIA8 is based on the Lund string fragmentation with different possible extensions. 
The~PYTHIA8 Monash~\cite{pythia_monash} is tuned on the LHC Run1 results, while the~additional junctions model~\cite{pythia_junctions} allow for more possibilities how the hadrons are created in the final state based on the colour indices of the constituent quarks. 
The~PYTHIA8 rope model~\cite{pythia_ropes} incorporates coherent interactions between charges  which allows for creation of colour ropes. 
Cluster hadronisation is used in HERWIG7~\cite{herwig}, where all hard processes and decays happen in the first stage and the free strings are subsequently clustered into hadrons.  
The EPOS~\cite{epos} generator implements the~core-corona model, where the dense core behaves hydrodynamicaly and it is surrounded by string decays in the corona. 
Data comparisons with the~different theoretical assumptions in these models can shed light on the actual hadronisation process preset in the~small collision systems.

\section{Event-topology dependent strangeness production}

One of the possible ways how to disentangle hard processes in an event, besides the jet finder algorithms, is usage of two-particle angular correlations. 
In this case, a leading charged primary particle is selected as a trigger particle to be a~proxy for the hard scattering. 
This is correlated with associated particles of interest, where, based on the differences in the~azimuthal angle and pseudorapidity, toward-leading and transverse-to-leading regions can be defined.
Integrating in these regions, particle production correlated with the~hard scattering (toward-leading region) and with the~soft processes (transverse-to-leading region) can be estimated. 

\begin{figure}
    \centering
    \includegraphics[width=0.48\linewidth]{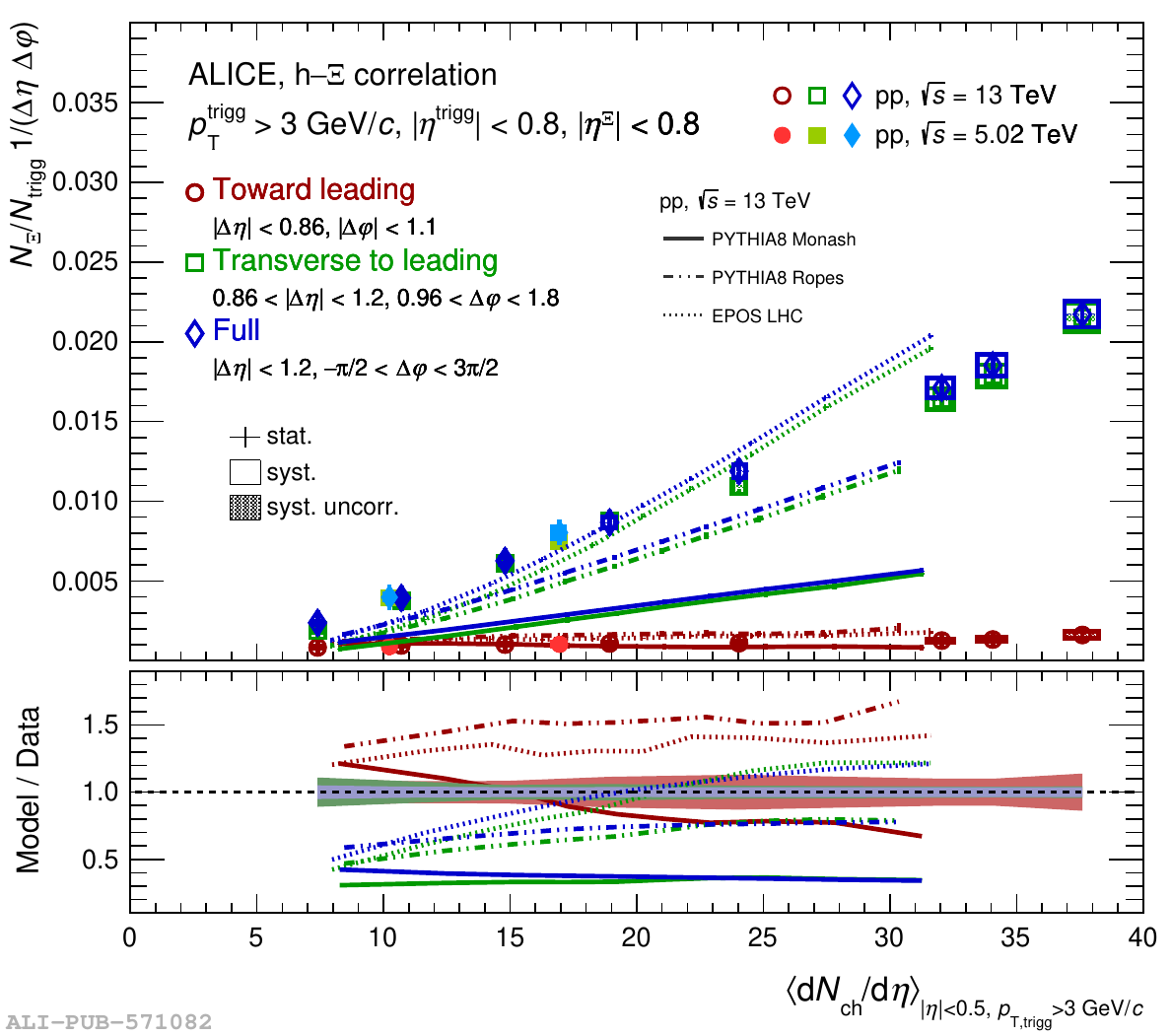}
    \includegraphics[width=0.48\linewidth]{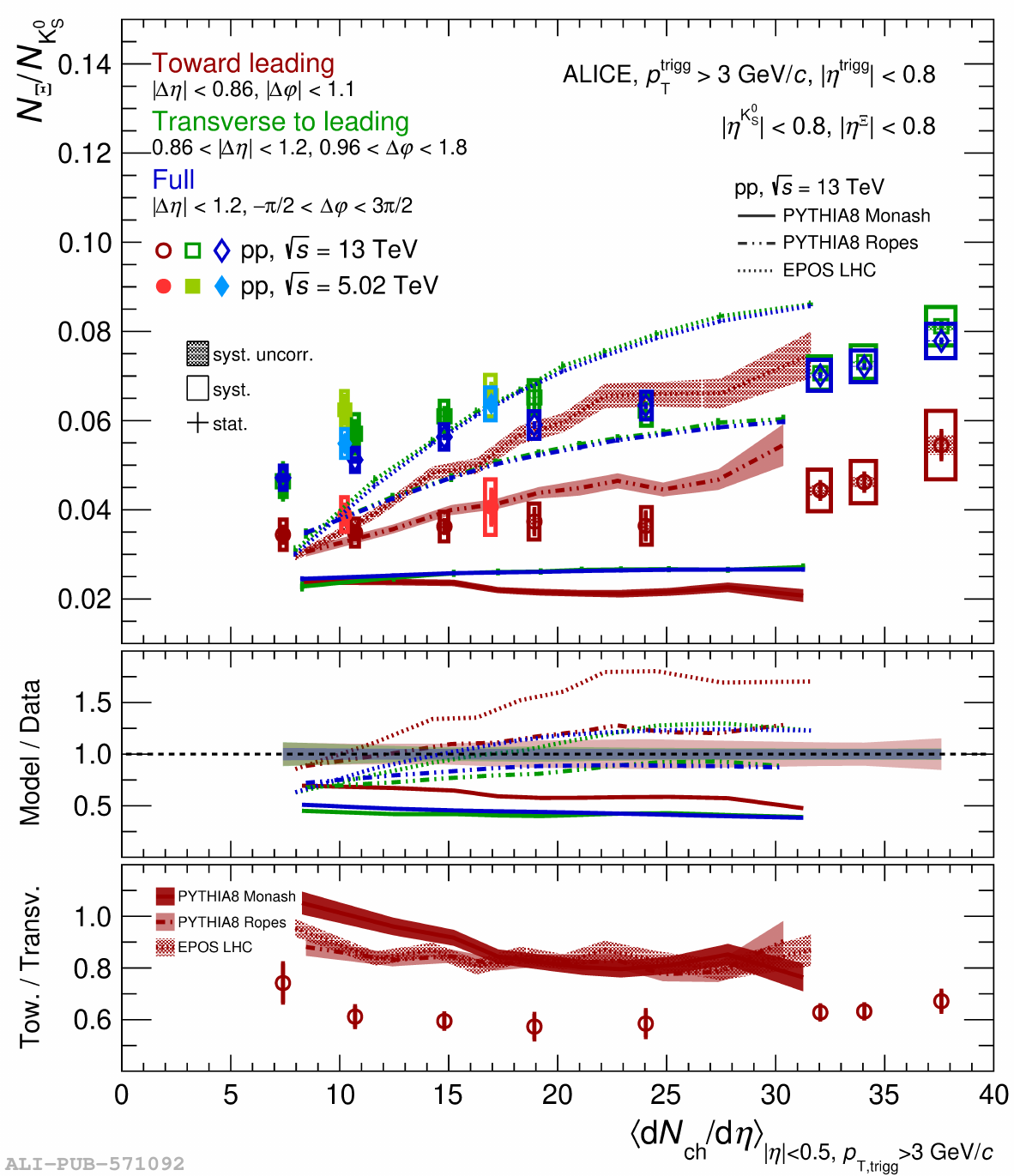}
    \caption{Left: Per-trigger $\Xi$ density as a function of charged particle multiplicity in the toward-leading, transverse-to-leading and full region. Right: $\Xi$/$\mathrm{K^0_S}$ ratio as a function with multiplicity~\cite{ALICE:corr}.}
    \label{fig:Xi_en}
\end{figure}

As shown in the~left panel of Fig.~\ref{fig:Xi_en}, the full $\Xi$ hyperon production is dominated by the~soft processes as the $\Xi$ density in the~full and transverse-to-leading region are mostly compatible, while the hard processes contribute only slightly to the total production. 
The natural increase of the total and transverse-to-leading $\Xi$ density is qualitatively described by the models. From a quantitative point of view, EPOS predicts even a steeper increase with multiplicity while both PYTHIA8 Monash and  PYTHIA8 with ropes underpredict the data. 
On the other hand, the toward leading production is overpredicted by both, EPOS LHC and PYTHIA8 Ropes. 

In order to study the strangeness enhancement, a ratio of $\Xi$ hyperon over $\mathrm{K^0_S}$ meson production is calculated in the same regions with respect to the leading particle and shown in the~right panel of Fig.~\ref{fig:Xi_en}. 
An~increase with multiplicity of this ratio is observed in all three studied regions.
Interestingly, the~steepness of the~increase is the~same for the~toward-leading and transverse-to-leading. 
This can be concluded from the~bottom panel of the~discussed figure, where a ratio of toward over transverse region is plotted and its shape is compatible with zero-degree polynomial. 
Based on this ratio, the~basic string fragmentation implemented in PYTHIA8 Monash can be ruled out, as it predicts the~$\Xi$/$\mathrm{K^0_S}$ to be constant as a function of multiplicity in all three regions, confirming that more processes are involved in the particle production in hadronic collisions.
One can see this by the~fact that both EPOS LHC and PYTHIA8 with ropes catch the~increase qualitatively, while the qualitative description needs to be improved. 

\section{Particle production mechanism}

\begin{figure}[ht!]
    \centering
    \includegraphics[width=0.32\linewidth]{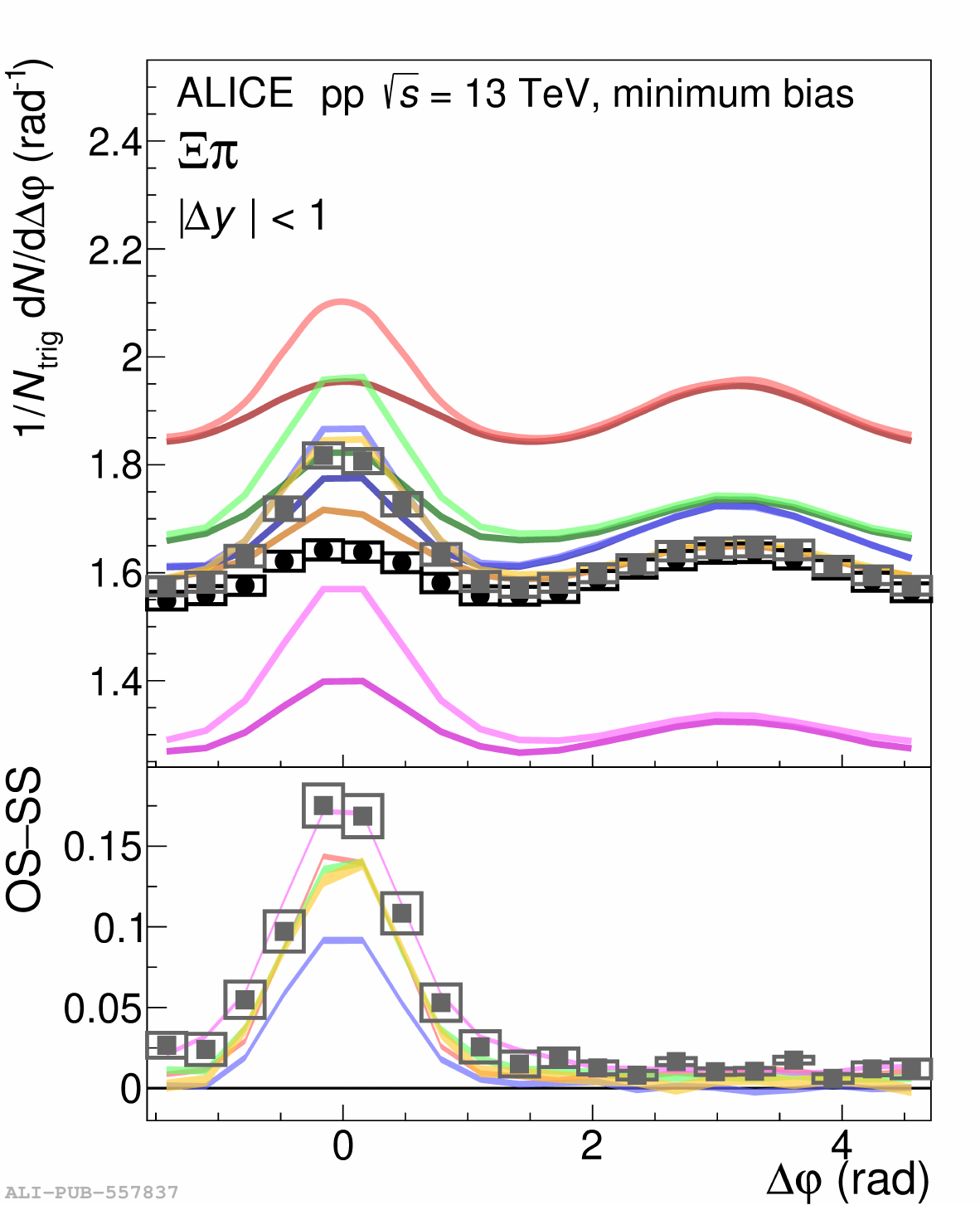}
    \includegraphics[width=0.32\linewidth]{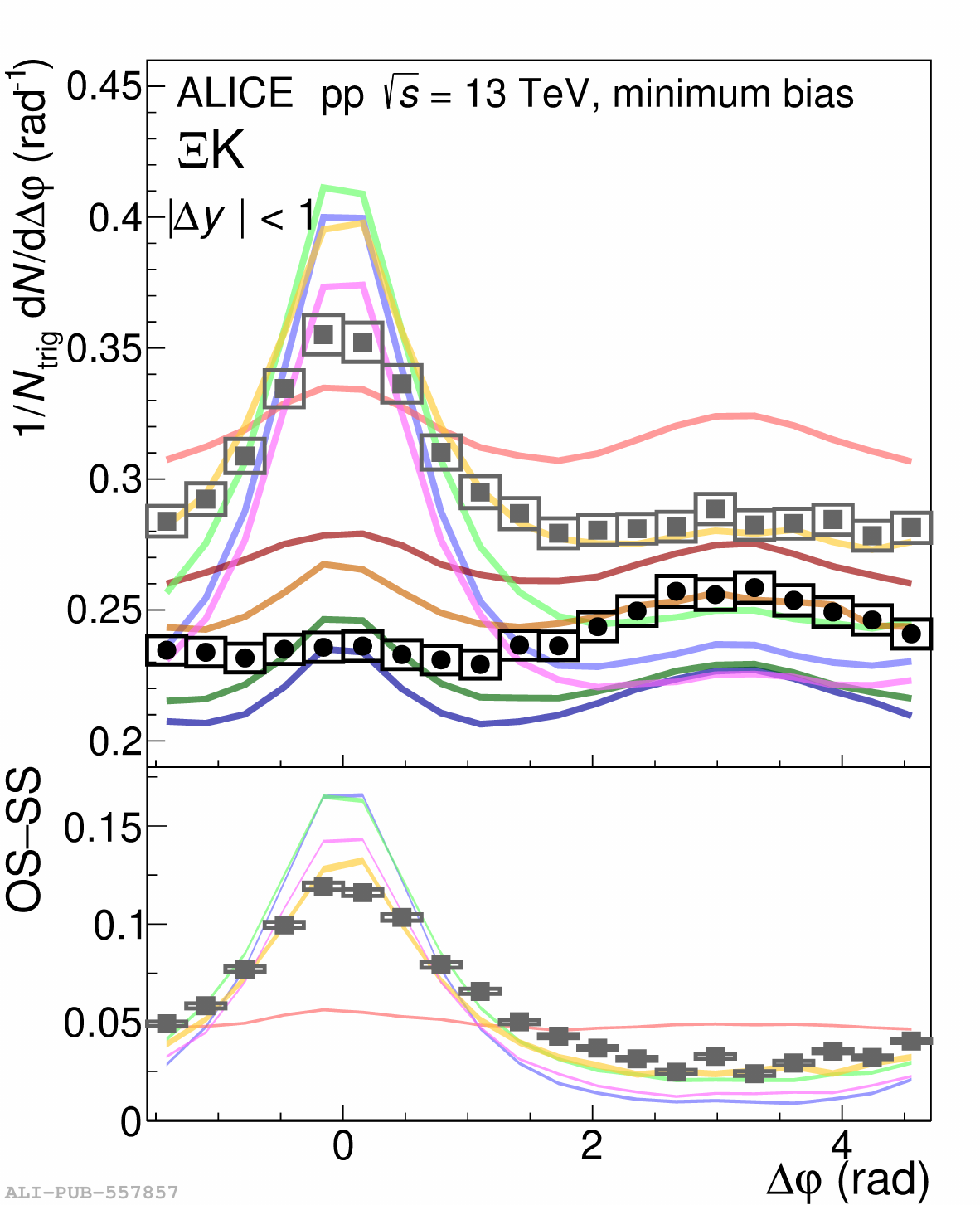}
    \caption{$\Xi \pi$ (left) and $\Xi$ K (right) correlation functions and the OS-SS differences (bottom panels) measured in pp collisions at 13~TeV compared with PYTHIA 8 Monash tune (violet), PYTHIA 8 with junctions enabled (green), PYTHIA 8 with junctions and ropes (yellow), EPOS-LHC (red), and HERWIG7 (pink)~\cite{ALICE:2023asw}. The darker shades and black points are showing SB correlation functions and the lighter ones OB correlation functions.}
    \label{fig:Xi_meson}
\end{figure}

\begin{figure}[h!]
    \centering
    \includegraphics[width=0.32\linewidth]{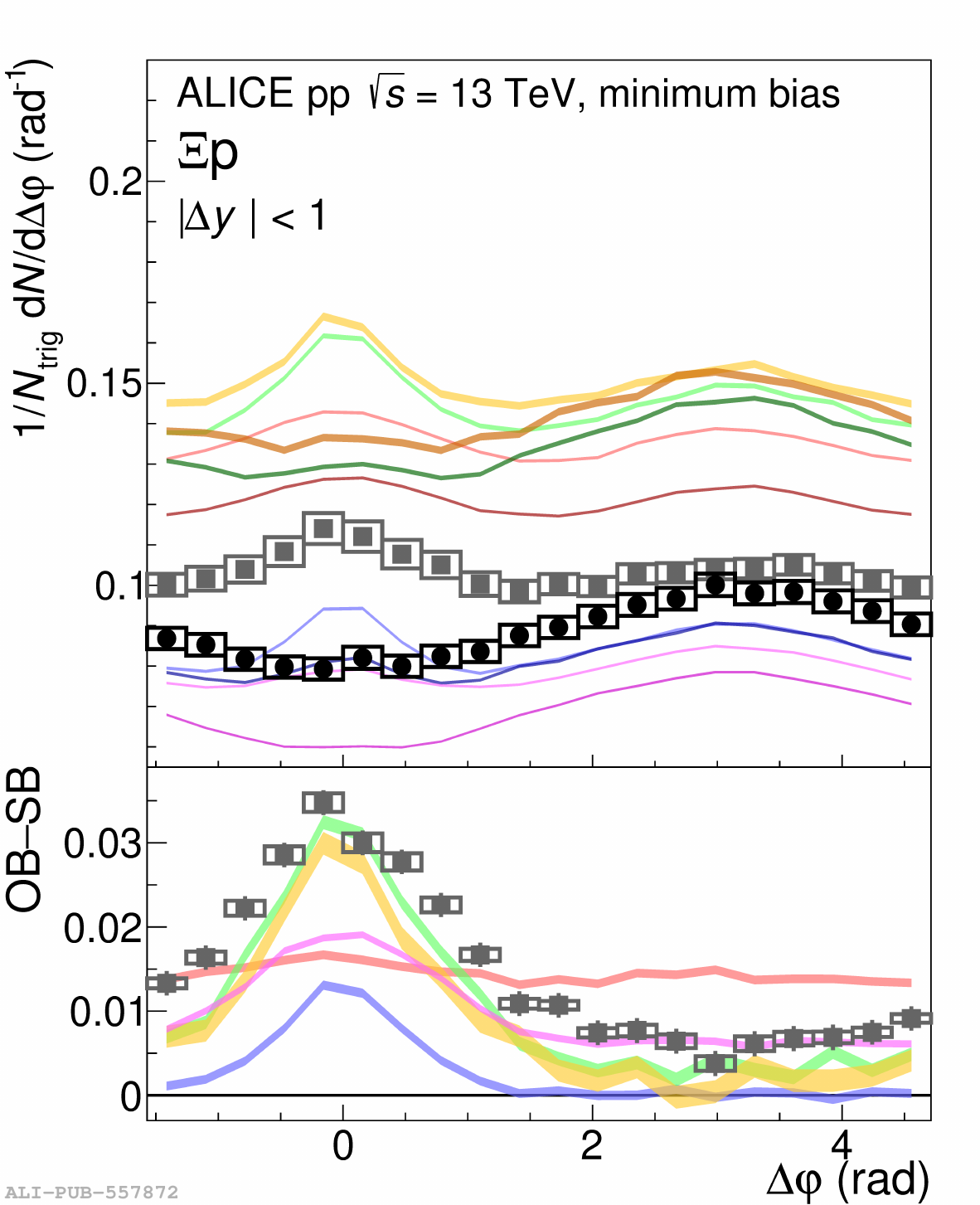}
    \includegraphics[width=0.32\linewidth]{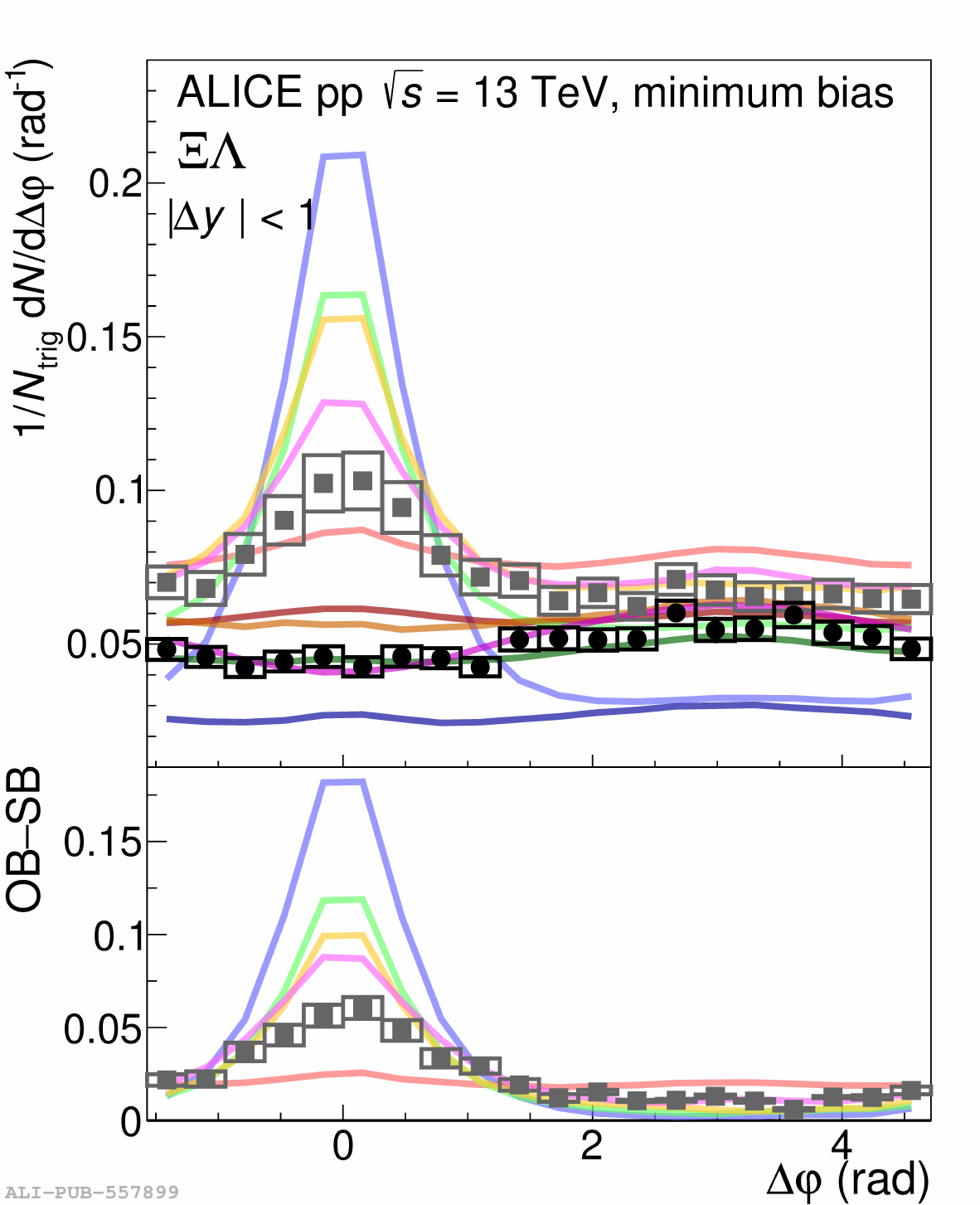}
    \includegraphics[width=0.32\linewidth]{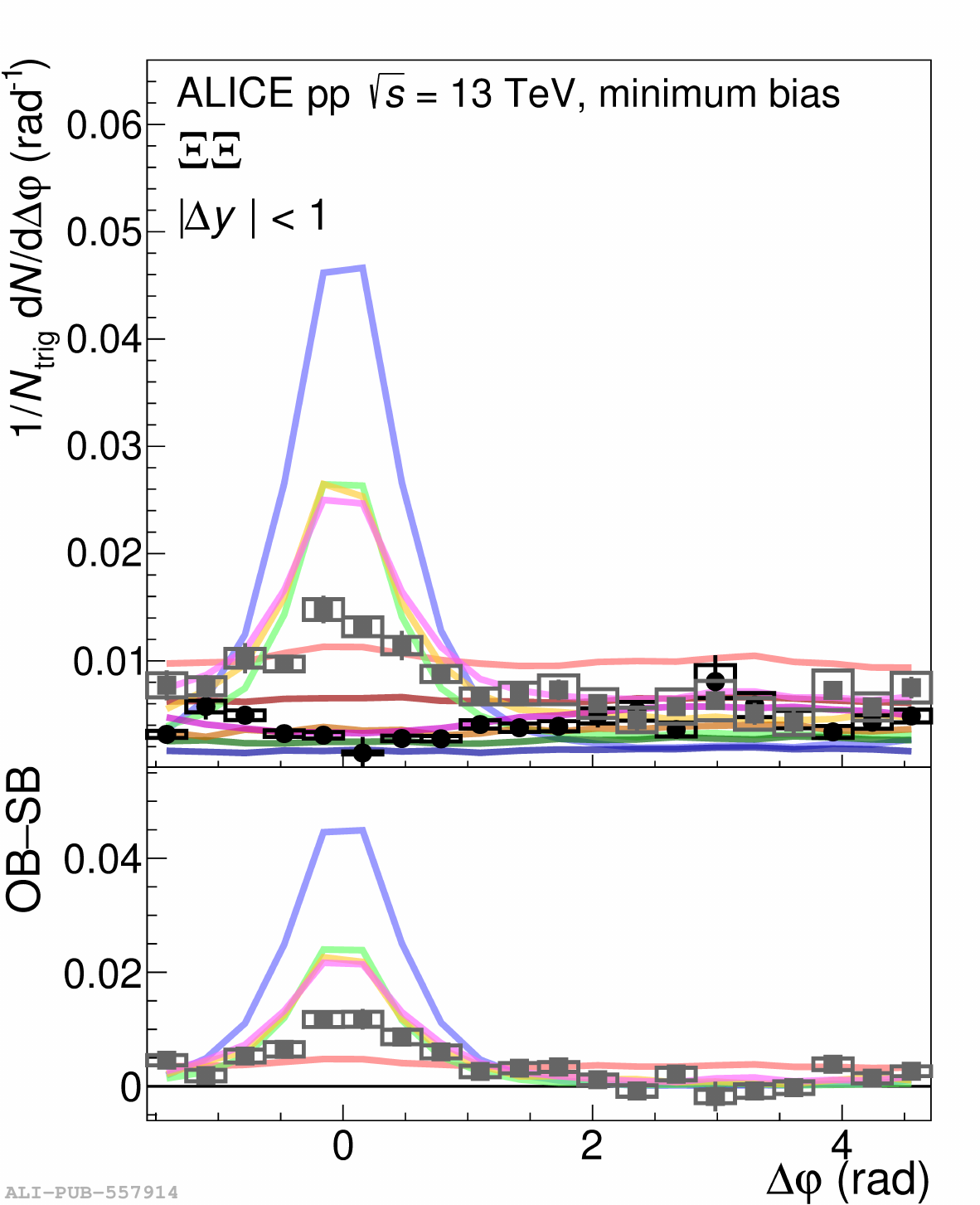}
    \caption{$\Xi p$ (left), $\Xi\Lambda$ (middle), and $\Xi\Xi$ (right) correlation functions and the OB-SB differences (bottom panels) measured in pp collisions at 13~TeV compared with PYTHIA8 Monash tune (violet), PYTHIA8 with junctions enabled (green), PYTHIA8 with junctions and ropes (yellow), EPOS LHC (red), and HERWIG7 (pink)~\cite{ALICE:2023asw}. The darker shades and black points are showing SB correlation functions and the lighter ones OB correlation functions.}
    \label{fig:Xi_baryon}
\end{figure}

Correlation studies between $\Xi$ hyperons and different identified hadrons can shed light into the~production mechanisms and help to understand how the~quantum numbers are conserved in the~phase space. 
A~proxy of the~balance function, a measure of distribution of balancing charges in momentum space, can be constructed by calculating a difference between opposite-sign~(OS),  (opposite-baryon-number~(OB)) and same-sign~(SS) (same-baryon-number~(SB)) correlation functions, where the uncorrelated part is subtracted. 
Such correlation functions between $\Xi$ baryon and $\pi$ and K mesons, and the difference OS-SS is shown in Fig.~\ref{fig:Xi_meson}.
From the comparison with the~different models can be seen that even though HERWIG7 can not catch the absolute magnitude of neither SS nor OS correlation functions, the balance of charge and strangeness is described correctly. 
In contrary, the core-corona model of EPOS LHC predicts a full decorrelation of strangeness that is not observed in the~data, while the charge balance is correctly described. 
All of the PYTHIA models are qualitatively in agreement with the~balance functions, but only the rope model can describe the magnitude of the correlation functions. 
By comparing the widths of the correlation function differences for pion and kaon, a wider peak can be observed for the $\Xi^\pm \mathrm{K^\pm}$ case~\cite{ALICE:2023asw}. 
The reason is not clear, but possible origins could be the quark mass difference or a partial diffusion of the $s$ quark.

Similar correlation functions for baryons (p, $\Lambda$, $\Xi$) and their OB-SB differences are shown in Fig.~\ref{fig:Xi_baryon} together with models.
While EPOS predicts full decorrelation of both strangeness and baryon number, PYTHIA8 Monash tune strongly overpredicts the short-range correlation of strangeness. 
Both considered extensions of PYTHIA8, with ropes and junctions, are able to describe the~balance of the baryon number, but overshoot slightly the short-range correlation of strangeness.
HERWIG7 is not able to fully describe the balance of neither the baryon number nor the~strangeness, but it is as the only model showing the near-side depletion in the SB correlation functions.
This is also observed in the pp+$\mathrm{\overline{p}\overline{p}}$ correlation function~\cite{ALICE:corrWUT}, but it has still an~unknown origin. 

\begin{figure}
    \centering
    \includegraphics[width=0.32\linewidth]{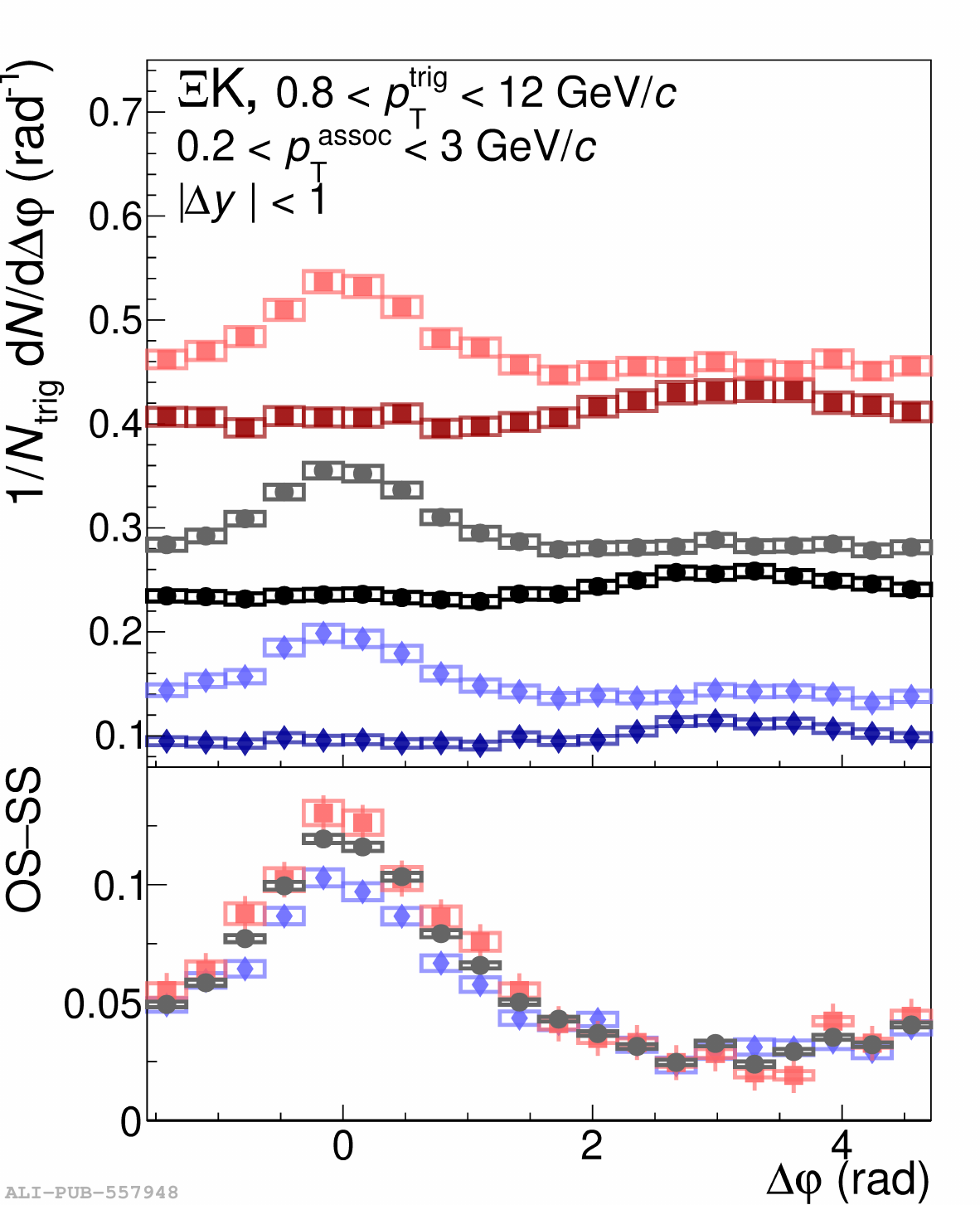}
    \includegraphics[width=0.32\linewidth]{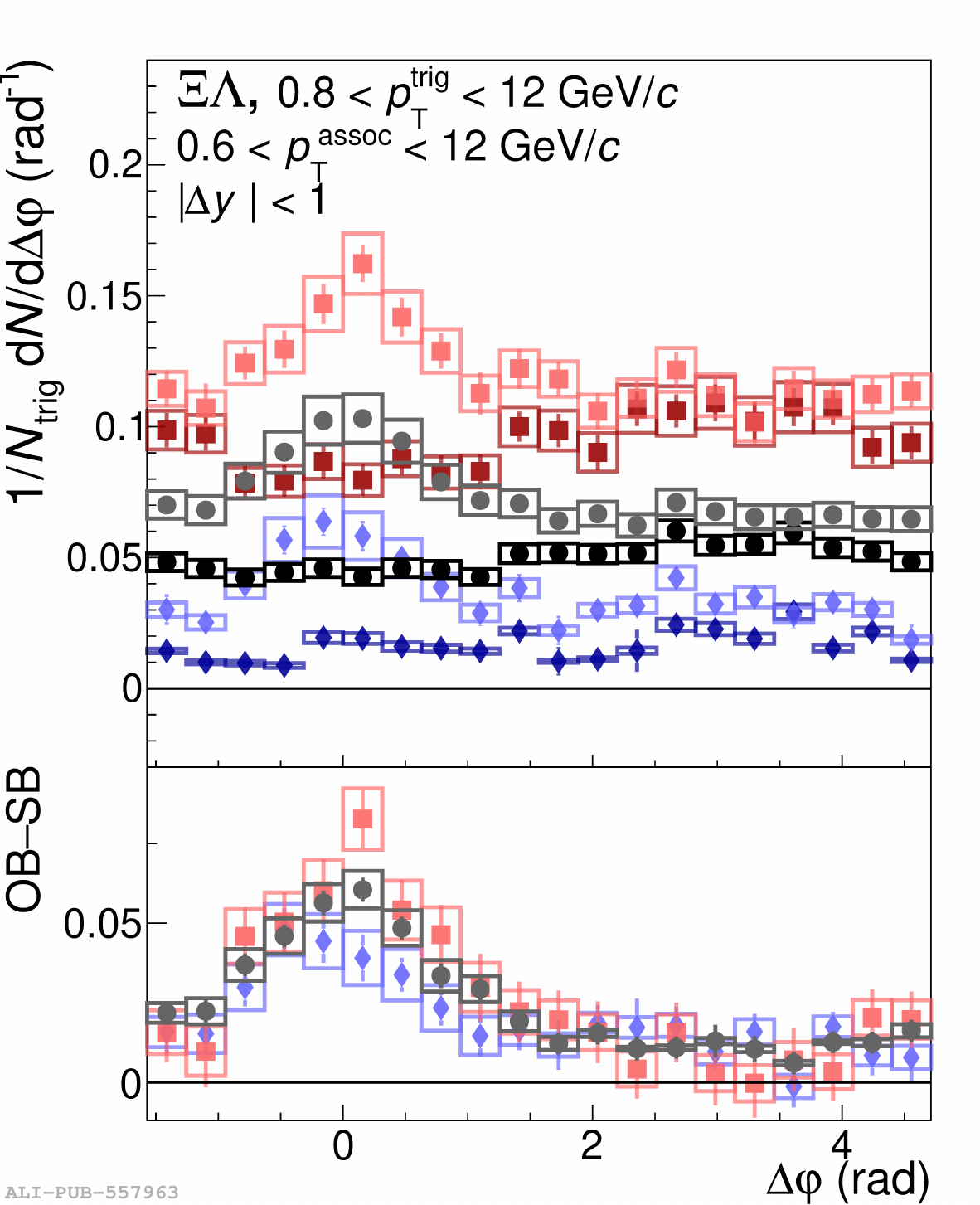}
    \includegraphics[width=0.32\linewidth]{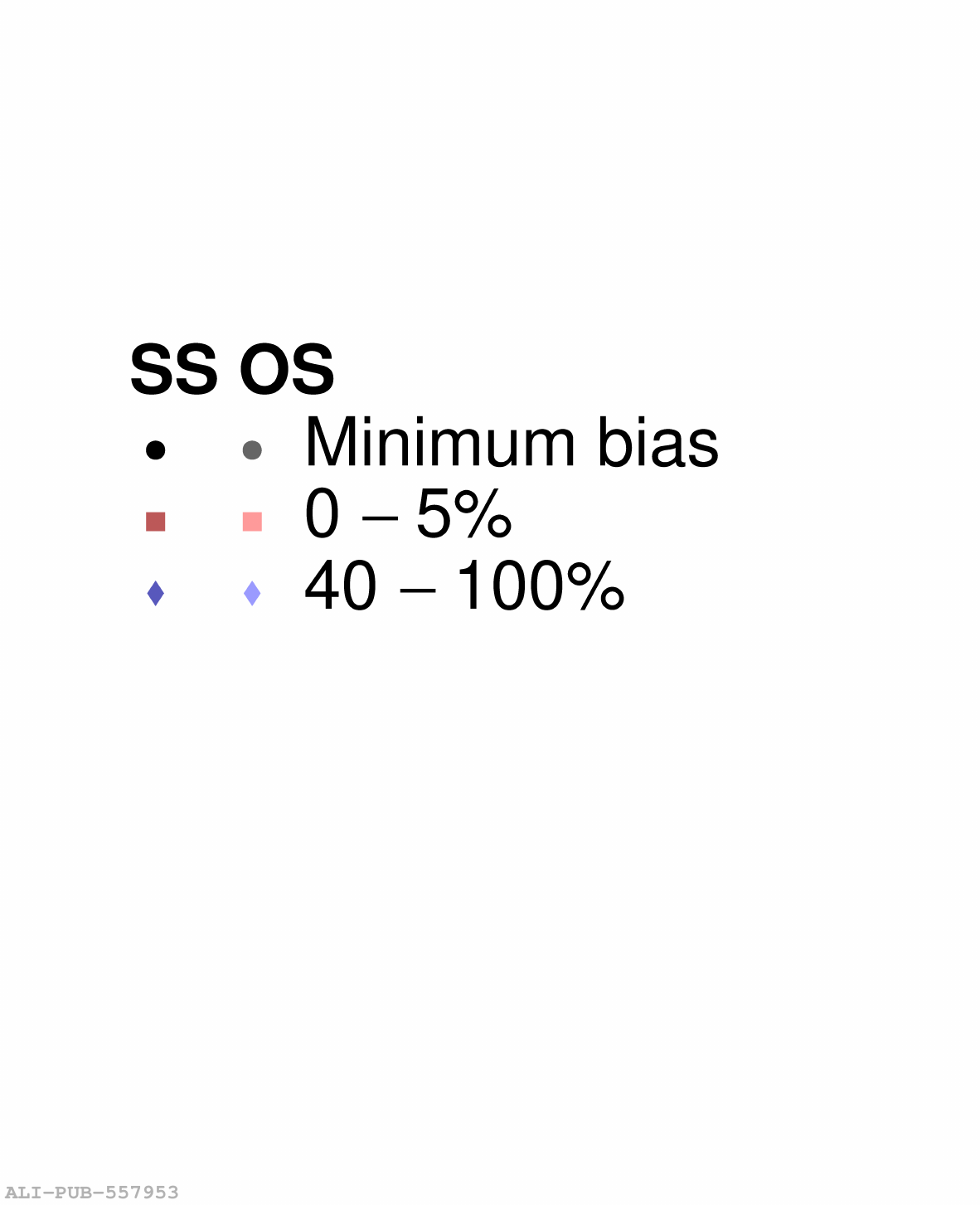}
    \caption{$\Xi$K (left) and $\Xi\Lambda$ (right) correlation functions and the OB-SB differences (bottom panels) measured in pp collisions at 13~TeV in different multiplicity classes~\cite{ALICE:2023asw}.}
    \label{fig:Xi_mult}
\end{figure}

Comparing the $\Xi$K and $\Xi\Lambda$ balance functions, it can be observed that strangeness is more~ balanced by mesons, where the correlated production increases with the collision multiplicity just slightly (Fig.~\ref{fig:Xi_mult}). 
Nevertheless, there is no evidence of multiplicity dependent particle production mechanism.


\section{Integrated strangeness production}
\begin{figure}[t!]
    \centering
    \includegraphics[width=0.48\linewidth]{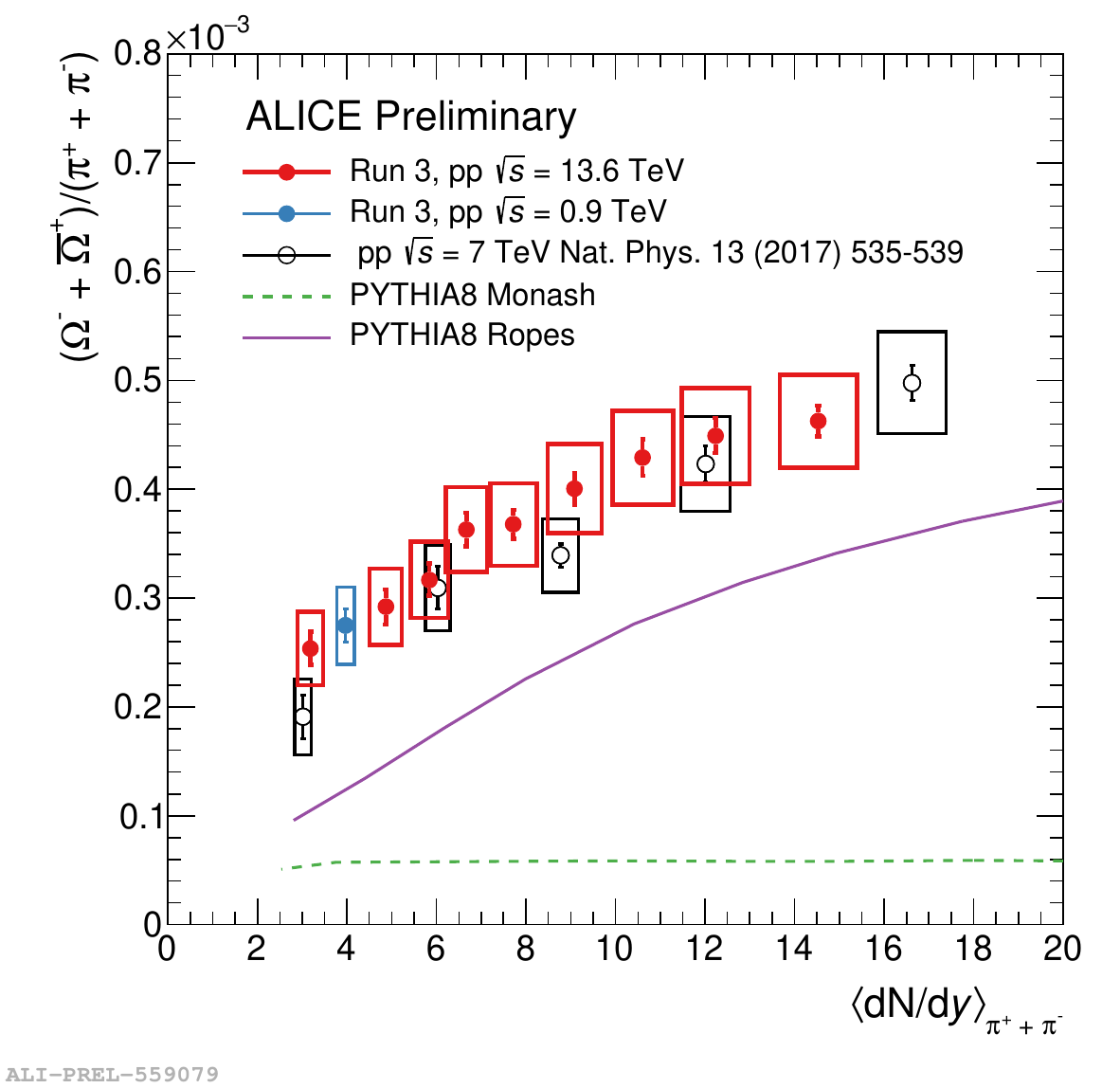}
    \caption{$\Omega$ baryon over pion ratio as function of pion multiplicity at midrapidity in pp collisions at 13.6~TeV and 0.9~TeV compared to the measurement at 7~TeV~\cite{ALICE:2016fzo} and with PYTHIA8 predictions. }
    \label{fig:Om_pion}
\end{figure}

A precise measurement of the integrated strangeness as a function of multiplicity in all collision systems is important to nail down any differences among the systems based on their size. 
The new $\Omega$ baryon over pion production measurements in pp collisions at 13.6~TeV and 900~GeV are shown in Fig.~\ref{fig:Om_pion} together with the measurement in pp collisions at 7~TeV. 
The results at all energies are aligned and show an increase with multiplicity which is qualitatively described by PYTHIA8 with rope hadronisation. 
With Run3 data-taking ongoing, the increasing statistics and software triggers will allow to extend the measurement to higher multiplicities in order to have a larger overlap region with peripheral p--Pb and Pb--Pb collisions.

\section{Conclusions}

Strangeness production measurements continue to bring new insights into the understanding of collision dynamics and particle production. 
New precise measurement of integrated $\Omega$ yield normalised to the~pion production is in agreement with the previous measurements showing enhanced production of strangeness with increasing multiplicity in pp collisions.
The correlation measurement shows that the integrated strangeness production is dominated by the transverse-to-leading particle production connected with soft processes. 
Nevertheless, increased production of strangeness is observed also in the~toward-leading region associated with the hard scattering. 
Measurement of correlation functions of $\Xi$ hyperon and identified hadrons points towards yet unclear multiplicity independent particle production mechanism. 
While the string breaking models overestimate the locally correlated $s\overline{s}$ production, it is strongly underestimated by models with thermalised medium.



\end{document}